\documentclass[aps,prl,reprint,groupedaddress,nofootinbib]{revtex4-1}

\usepackage{graphicx}
\usepackage{amsmath}
\usepackage{bm}
\usepackage{dsfont}
\usepackage{epsfig}
\usepackage[usenames]{color}
\usepackage{comment}

\def\TBCOMMENT#1{}
\def\COMMENT#1{}

\def\ms{\mathrm{ms}}

\def\deg{^\circ}
\def\persec{\mathrm{s^{-1}}}
\def\mm{\mathrm{mm}}

\def\rhodot{\dot{\rho}}
\def\Tcorr90{T_{\mathrm{c}}}

\def\DeltaRhoMax{\Delta\rho}

\def\phib{\phi_\mathrm{b}}
\def\thetab{\theta_\mathrm{b}}

\def\xb{\mathbf{\hat{x}}_\mathrm{b}}
\def\yb{\mathbf{\hat{y}}_\mathrm{b}}
\def\zb{\mathbf{\hat{z}}_\mathrm{b}}

\begin{document}

\title{Roll control in fruit flies}




\author{Tsevi Beatus$^1$ John M. Guckenheimer$^2$, and Itai Cohen$^1$}
\affiliation{Departments of $^1$Physics and $^2$Mathematics, Cornell University, Ithaca, New York 14853, USA}

\date{\today}

\begin{abstract}
Due to aerodynamic instabilities, stabilizing flapping flight requires ever-present fast corrective actions. Here we investigate how flies control body roll angle, their most susceptible degree of freedom. We glue a magnet to each fly, apply a short magnetic pulse that rolls it in mid-air, and film the corrective maneuver. Flies correct perturbations of up to $100\deg$ within $30\pm7\ms$ by applying a stroke-amplitude asymmetry that is well described by a linear PI controller. The response latency is $\sim5\ms$, making the roll correction reflex one of the fastest in the animal kingdom. 
\end{abstract}



\maketitle

Locomoting organisms evolved mechanisms to control their motion and maintain stability against mechanical disturbances. The control challenge is prominent for small flying insects since their small moment of inertia renders them susceptible even to gentle air currents \cite{CombesTurbulencePNAS2009, RaviCombesJEB2013, OrtegaHedrickJEB2013, VanceHumbert2013}. Moreover, they fly at intermediate Reynolds numbers $\textrm{Re}=10^2-10^4$, in which flows are unsteady \cite{SaneReview2003, WangReview2005}. Most importantly, flapping flight is aerodynamically unstable, on a time scale of a few wing-beats \cite{TaylorThomasJEB2003, SunBumblebeeJEB2005, TaylorZibkowski2005, SunDynamicStability2007, LiuCFD2010, FaruqueHumbert2010a, RistrophPitchInterface2013,  GaoCFD2011, LiangSunInterface2013, XuSunCFD2013, ZhangSunLateral2010}. It is, therefore, intriguing how insects overcome such control challenges and manage to fly with impressive stability, maneuverability and robustness, outmaneuvering any man-made flying device. 
 
Among the body Euler angles -- yaw, pitch and roll -- roll is most sensitive to perturbing torques since the moment of inertia along the insect's long axis is smallest \cite{CombesTurbulencePNAS2009, RaviCombesJEB2013}. Recent fluid dynamics simulations suggest this degree of freedom is unstable due to an unsteady aerodynamic mechanism, where roll is positively coupled to sideways motion via asymmetry of the leading-edge vortex attached to each wing \cite{ZhangSunLateral2010, LiangSunInterface2013, XuSunCFD2013, GaoCFD2011}. Such results indicate flies can lose their body attitude due to roll perturbations within $4$ wing-beats. Controlling roll is also crucial for maintaining direction and altitude. Thus, any basic understanding of insect flight demands quantitative analysis of roll control. 

Previous studies used tethered animals to measure changes in wing motion in response to imposed roll rotations \cite{ NalbachNonOrthogonal1994, DickinsonHaltere1999, ShermanDickinsonJEB2003, ShermanDickinsonJEB2004, SugiuraDickinsonPlotOne2009, HengstenbergRoll1986} and visual roll stimuli \cite{SrinivasanVisual1977, HengstenbergRoll1986, WaldmannLocust1988, LehmannDickinson1998, ZankerIIIControl1990, SugiuraDickinsonPlotOne2009, WindsorInterface2014}. In such experiments, however, the tethered insect does not control its motion and often exhibits wing kinematics and torques qualitatively different from those in free-flight \cite{BenderDickinsonJEB2006}. More recently, free-flight experiments used turbulent wakes \cite{CombesTurbulencePNAS2009, RaviCombesJEB2013, OrtegaHedrickJEB2013} and impulsive gusts \cite{VanceHumbert2013} to perturb insects, highlighting the sensitivity of their roll angle to perturbations. Understanding the roll control mechanism however, requires fast and accurate quantitative measurements of wing and body kinematics in response to controlled mid-air perturbation impulses -- a methodology that was recently applied to study yaw control \cite{RistrophPNAS2010}.

Here, we perturb a fruit-fly (\emph{Drosophila melanogaster}) by gluing a magnet to its back and applying a $\sim5\ms$ magnetic pulse that rolls it in mid-air. We use high speed video to film the fly's corrective maneuver and measure its wing and body kinematics \cite{RistrophHRMT_JEB2009}. We find that flies manage to correct for roll perturbations of up to $100\deg$ within $6.8\pm1.6$ wing-beats ($30\pm7\ms$). Moreover, the active correction is nearly complete by the time the visual system can trigger a response \cite{FotowatEscapeResponse2009, LandCollet1974}. The flies generate corrective torques by applying a stroke-amplitude asymmetry that can be described by the output of a linear PI controller. The asymmetry starts only one wing-beat ($\sim5\ms$) after the onset of the perturbation, making the roll correction reflex one of the fastest in the animal kingdom \cite{EatonBook1984}. 

\begin{figure} 
\includegraphics[scale=0.90]{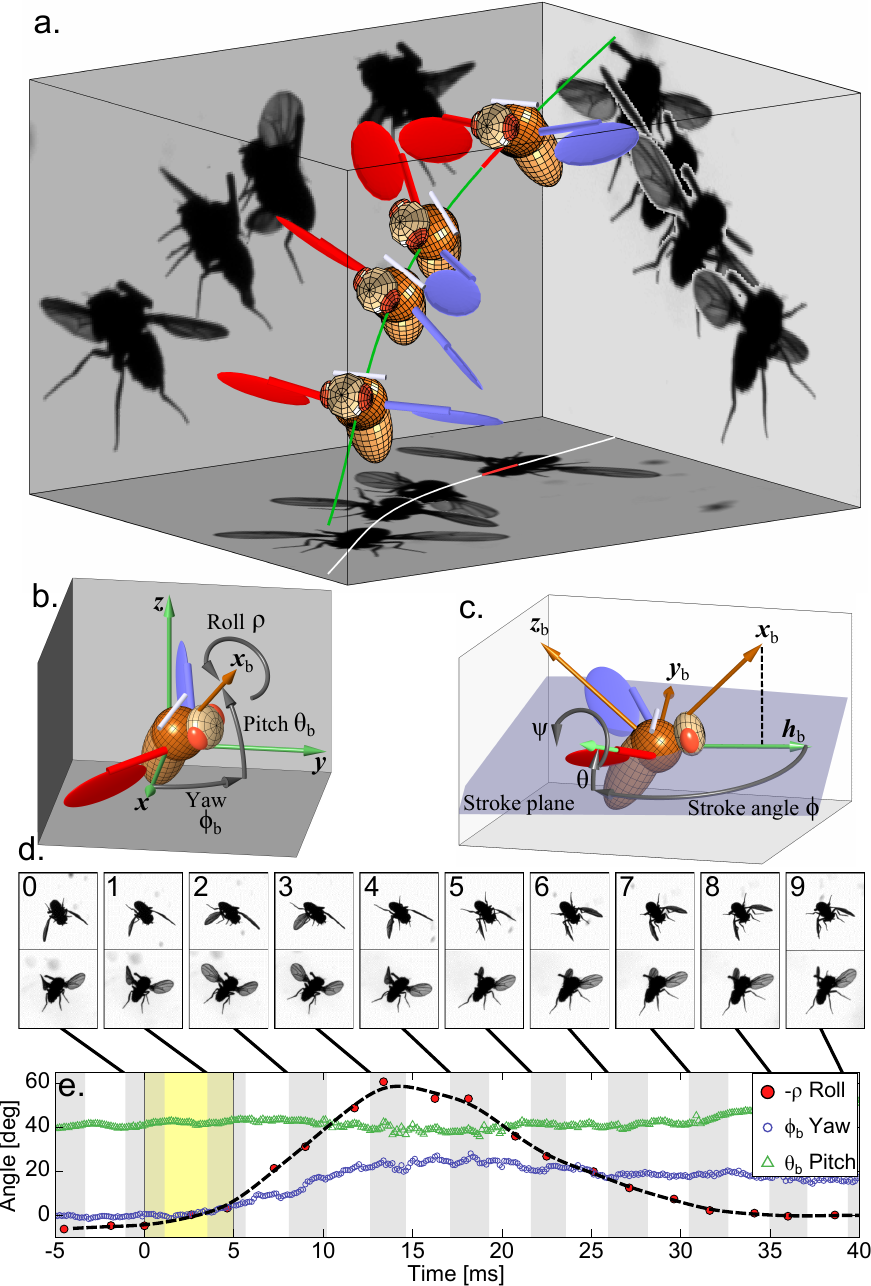}
\caption{\label{Fig1:TheSystem}Roll perturbation and correction: (a) Images from 3 orthogonal cameras of a fly undergoing a roll perturbation and correction maneuver. Each panel shows $4$ superposed images. The 3D-rendered fly represents the kinematic data of the body and wings. The perturbation location (red line) is shown on the fly's center-of-mass trajectory (green). In the second snapshot the fly is rolled $60\deg$ to its left. (b) Definition of the body Euler angles with respect to the lab frame. $\xb$ is the long body axis. (c) Definition of Euler angles and body frame $(\xb, \yb, \zb)$. Wing angles are measured in the body frame with respect to the stroke plane (shaded blue, see SI). (d) Top and side-view snapshots of 10 consecutive wing-strokes of the maneuver. Snapshots were taken at the stroke phase where the wings are at their forward-most position. The perturbation wing-beat is numbered $0$. (e) The body Euler angles during the maneuver. The perturbation torque was on between $0-5\ms$ (yellow stripe). The white and gray stripes represent forward- and back-strokes, respectively. Yaw and pitch were sampled at $8000\mathrm{Hz}$. Roll was measured manually in the middle of each half-stroke and smoothed by a spline (dashed line). Measurement errors are comparable to the symbols size.} 
\end{figure}

\emph{The experimental system:} To exert mid-air mechanical perturbations along roll we glue a magnetic pin, $1.5-2\mm$ long, to the dorsal thoracic surface of each fly (Fig.~1). The pin is oriented horizontally and perpendicular to the body axis. In each experiment $\sim 15$ flies are released in a transparent chamber equipped with two Helmholtz coils that are used to generate a vertical magnetic field ($\sim10^{-2}\mathrm{Tesla}$). When a fly crosses the filming volume, a laser-trigger initiates video recording at $8000 \: \mathrm{frames}$ $\persec$ along three orthogonal axes, as well as a magnetic pulse lasting $5\ms$, or $1$ wing beat \cite{RistrophHRMT_JEB2009, RistrophPNAS2010}. Since fruit flies fly with their body axis pitched up at $\sim45\deg$ and since the moment of inertia along their body axis is $\sim4$ times smaller than the other axes, the largest deflection is generated along the body roll axis, with smaller perturbations along pitch and yaw (Fig.~1a,b). Using a custom image analysis algorithm \cite{RistrophHRMT_JEB2009}, we extract a 3D kinematic description of the fly (Fig.~1) consisting of its body position and orientation (Fig.~1b) as well as the Euler angles (Fig.~1c) for both wings (see Supplementary Information (SI)). We analyzed $20$ sequences that span a perturbation range between $5\deg$ and $100\deg$, in which the flies perform a steady flight before and after the correction maneuver.

\emph{Roll correction mechanism:} A representative example of a fly recovering from a $60\deg$ roll perturbation is shown in Fig.~1 and Movie~1. Top and side-view snapshots of consecutive wing-strokes for the maneuver are shown in Fig.~1d. The images correspond to the phase in the stroke where the wings are in their forward-most position. The body Euler angles, roll ($\rho$), yaw ($\phib$) and pitch ($\thetab$) are plotted in Fig.~1e. The magnetic field was applied between $t=0-5\ms$ (yellow stripe) and induced a maximum roll velocity of $7000\deg\persec$ resulting in a deflection of $60\deg$ within $t=13.5\ms$ (wing-beat no. 3 in Fig 1d). The fly recovered its initial roll angle within $35\ms$, or $8$ wing-beats. The top view images show a clear asymmetry in wing stroke angles during the maneuver, which starts a single wing-beat ($5\ms$) after the perturbation (Fig.~1d, frames $1-4$). During the maneuver the left wing stroke amplitude increases while the right wing stroke amplitude decreases. The fly also spreads its legs from their folded flight position (frames $4-8$) as in a typical landing response \cite{goodman1960landing, borst1986time, vanBreugel2012visual}. In addition to the roll perturbation, smaller deflections of $25\deg$ left in yaw and $5\deg$ down in pitch were also induced, since the applied torque is not completely aligned with a principal body axis (Fig. 1e, SI).

\begin{figure}
\includegraphics[scale=1]{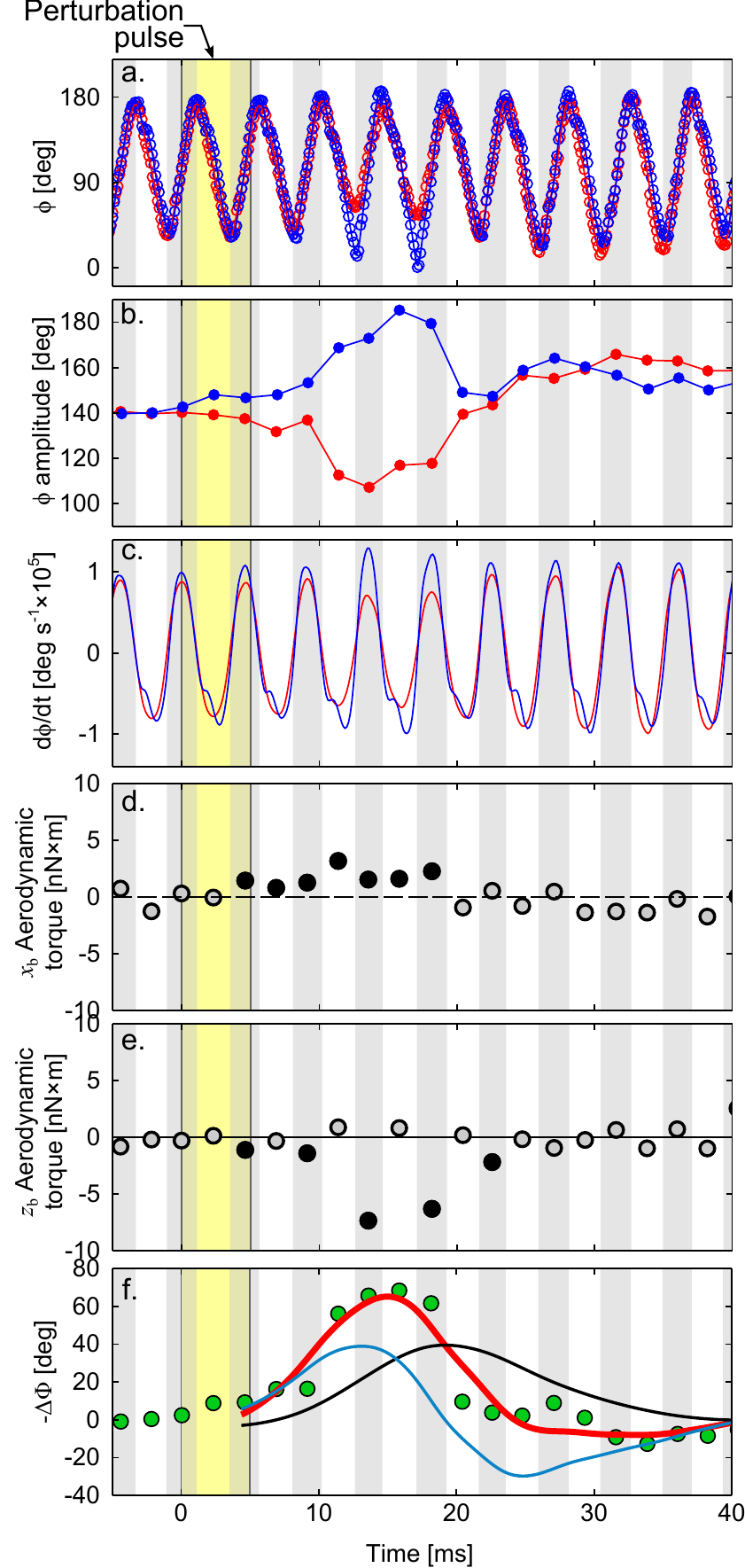}
\caption{\label{Fig2:CorrectionMechanism}Roll correction mechanism for the correction maneuver shown in Fig.~1. (a-c) Wing stroke kinematics versus time: (a) The stroke angle $\phi$ of the right (red) and left (blue) wings; (b) their peak-to-peak amplitude $\Phi$, and (c) their angular velocity $\dot{\phi}$ (c). (d-e) Mean aerodynamic torque along each half stroke, calculated from the measured wing kinematics using a quasi-steady state aerodynamic force model. Solid symbols highlight the correcting wing strokes. (d) the torque component along the body axis $\xb$. Positive torque induces a corrective right roll; (e) The torque component along $\zb$. Negative values have corrective effect. (f) Wing stroke amplitude difference $\Delta\Phi$ (green), and a fit for a PI controller (eq. 1, red), with $\Delta T=4.4\pm1\ms$, $K_\mathrm{p}=6\pm2\ms$, and $K_\mathrm{i}=0.7\pm0.2$. The contributions of the $1^\textrm{st}$ and $2^\textrm{nd}$ terms of Eq. 1 are shown in blue and black, respectively. Measurement errors in (a-f) are comparable to the symbols size.}
\end{figure}

We quantify the asymmetry in wing kinematics by plotting the stroke angles for the left (blue) and right (red) wings during the maneuver (Fig.~2a). As shown in Fig.~2b, we find large differences -- up to $70\deg$ -- between their peak-to-peak amplitudes (see SI). The amplitude asymmetry began only $1$ wing-beat after the onset of the perturbation, and lasted for $5$ wing-beats. The flapping frequency of both wings remained nearly constant during the maneuver. Hence, to maintain the amplitude asymmetry, the right wing moved faster than the left (Fig.~2c). To first order, this difference in velocity leads to an asymmetry in the aerodynamic forces produced by the two wings and generates a correcting torque. 

To calculate the aerodynamic torque generated by the insect, we used the measured wing and body kinematics combined with a quasi-steady-state model for the aerodynamic force \cite{DickinsonScience1999} produced by each wing (SI).  The calculated torques are similar for other quasi-steady-state force models \cite{SaneDickinsonJEB2002, BermanWangJFM2007} as well. The components of the aerodynamic torque vector along the $\xb$ and $\zb$ body axes were averaged over half-strokes and plotted in Fig.~2d,e  (see Fig.~1c for axes definition). The torque magnitude is roughly $5\,\textrm{nN}\cdot\textrm{m}$ and is comparable to torques exerted by tethered fruit flies \cite{SugiuraDickinsonPlotOne2009}. Both the $\xb$ and $\zb$ torque components have a corrective effect along roll, and both exhibit distinct peaks (solid circles) that appear simultaneously with the stroke-amplitude asymmetry. 

This amplitude asymmetry can be described by the response of a linear, proportional-integral (PI) controller:
\begin{equation}
\label{Eq:Controller} 
\Delta\Phi(t) = K_\mathrm{p} \dot{\rho}(t-\Delta T) + K_\mathrm{i} \rho(t-\Delta T).
\end{equation}
Here, the output $\Delta\Phi$ is the difference between the right and left wing stroke amplitudes, and the controller's input is the body roll velocity, $\rhodot$, which flies measure using their gyroscopic sensor system associated with the haltere organs \cite{Pringle1948, NalbachHalteres1993, DickinsonHaltere1999}. The controller is defined by three parameters: the proportional gain $K_\mathrm{p}$, the integral gain $K_\mathrm{i}$, and a delay $\Delta T$ that describes the neuro-muscular response time. Fitting these three parameters using the measured $\rhodot$, $\rho$ and $\Delta\Phi$ (Fig.~2f), we find this controller response (red curve) is sufficient to reproduce the time-dependence of $\Delta\Phi$ (green circles). Moreover, the fast rise time can be attributed to the term proportional to the roll velocity (blue curve). Simpler models, such as I- and P-controllers, cannot reproduce the response as well (SI). As such, this PI model is the simplest continuous linear control mechanism consistent with our observations. 

The salient feature of the correction mechanism -- a wing stroke amplitude asymmetry that starts $\sim1$ wing-beat after the perturbation onset and lasts for several wing-beats -- was observed in all our recorded perturbation events (Fig.~3a). Moreover, this feature was robust to variability in initial flight pose and velocity. For example, the same mechanism was observed for hovering flies (Movie 2), flies with nonzero roll angle at the perturbation onset (Movie 3), and even flies subject to two consecutive perturbing pulses (Movie 4, Figs. S2, S3). In-depth analysis of $11$ correction maneuvers showed the $\Delta\Phi$ response is consistent with the PI controller model. Fitting the control parameters for each maneuver separately, we find $K_\mathrm{p}=4.8\pm2.4\ms$ and $K_\mathrm{i}=0.6\pm0.3$ (mean $\pm$ standard deviation). The mean response time $\Delta T=4.6\pm1\ms$ is comparable to a single wing-beat period (Fig.~3a). This response time is $3.5$ times faster than the flies' response to yaw perturbations \cite{RistrophPNAS2010} and $2.5$ times faster than their response to pitch perturbations \cite{RistrophPitchInterface2013}, underscoring the importance of roll control for flight.

\begin{figure}
\includegraphics[scale=0.78]{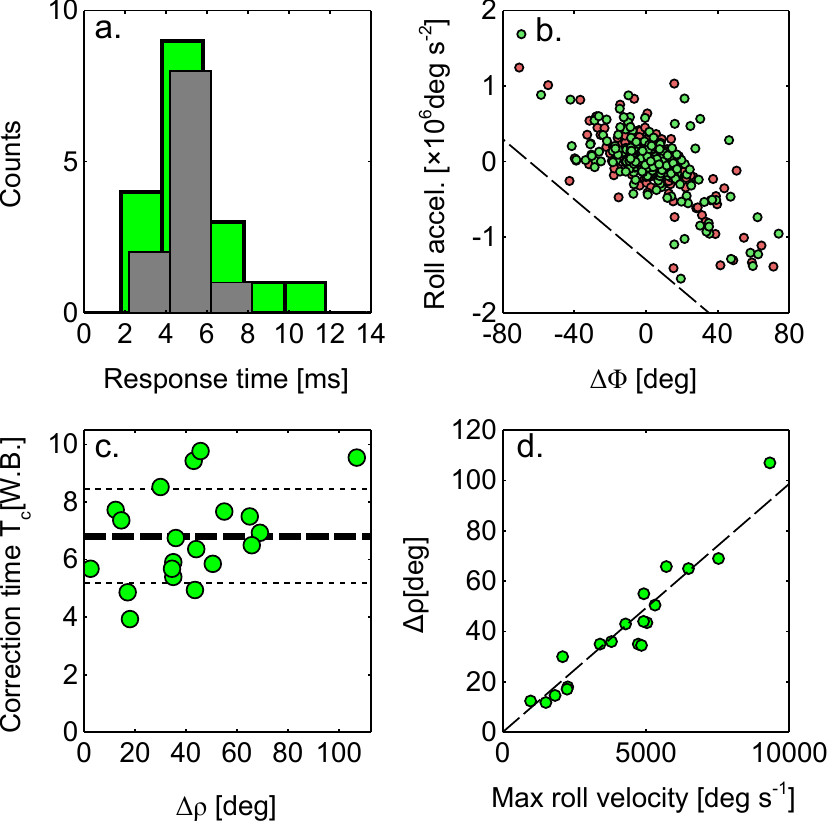}
\caption{\label{Fig3}(a) Response time histograms for multiple perturbation events: Delay time of fitted PI controllers (dark gray) and the latency time to reach $\Delta\Phi=10\deg$ measured from the wing kinematics (green). Histograms are artificially shifted to improve visibility. (b) Mean roll acceleration as a function of $\Delta\Phi$ for $299$ forward strokes (red) and $304$ back strokes (green). The correlation coefficient of these two quantities is $-0.67$ ($p \textrm{-value} <10^{-12}$). The dashed line has a slope of $-20\cdot 10^3\mathrm{s^{-2}}$. (c) Roll correction time $\Tcorr90$ measured in wing-beats as a function of $\DeltaRhoMax$ for multiple events. The mean correction time (thick dashed line) is $6.8\pm1.6$ wing-beats (mean$\pm$standard deviation). Thin dashed lines indicate $1\sigma$ margins. (d)  Maximum roll displacement $\DeltaRhoMax$ as a function of the maximum roll angular velocity for multiple events. The dashed line has a fitted slope of $9.9\ms$ with a $95\%$ confidence interval of $\pm0.75\ms$. }
\end{figure}

The torques generated by the wing asymmetry (Fig.~2d,e) are qualitatively correcting. However, quantitatively relating the asymmetric wing kinematics to the 3D rotational body dynamics is difficult due to noise in the measurements and unsteady aerodynamic effects. To further illustrate that the wing asymmetry indeed generates corrective roll dynamics, we determine the mean roll acceleration for $\sim600$ half-strokes and plot it as a function of the measured $\Delta\Phi$ (Fig.~3b). We find that the $\Delta\Phi$ asymmetry is indeed negatively correlated with the roll acceleration generated by the fly.  Thus, for example, negative $\Delta\Phi$ is correlated with positive roll acceleration, as in the measurement shown in Figs. 1, 2, S2 and S3. 

We also find that the data show two hallmarks of linear control. First, the correction time is insensitive to the maximum roll deflection, $\DeltaRhoMax$. Here the correction time $\Tcorr90$ is defined as the time between the onset of the perturbation and the moment at which the roll angle reaches $10\%$ of $\DeltaRhoMax$. Plotting $\Tcorr90$ as a function of $\DeltaRhoMax$ shows the correction time is $6.8 \pm 1.6$ wing beats (mean $\pm$ s.d., $n=20$) with little dependence on the perturbation amplitude (Fig.~3c). Second, we find that $\DeltaRhoMax$ increases linearly with the maximum roll velocity (Fig.~3d). Collectively, these data suggest that as with yaw \cite{RistrophPNAS2010}, the flies' response to roll perturbations can be effectively described by a reduced order model consisting of a linear PI controller with time delay.

\emph{Extreme perturbations:} To test the linear control model we challenged the flies with extreme perturbations in which they were spun multiple times in mid air by a series of magnetic pulses. The fly shown in Fig.~4a and Movie 5 was rotated $8$ times to its right. The accumulated roll angle exceeded $3000\deg$ (Fig.~4b) and the maximum roll velocity was over $60,000\deg\persec$ (Fig.~4c). During the perturbation, the fly was unable to oppose the magnetic torque. In fact, the right wing, which in a typical correction maneuver flaps with a larger stroke amplitude, hardly flapped at all and occasionally seemed disconnected from its flight power muscles. We captured 3 such events, all showing the same behavior. Remarkably, once the magnetic pulses stopped, the flies regained control within $3-4$ wing-beats. Our calculations show the roll deceleration is only explained by active flapping, rather than passive damping due to the wings (SI).
\begin{figure}
\includegraphics[scale=0.95]{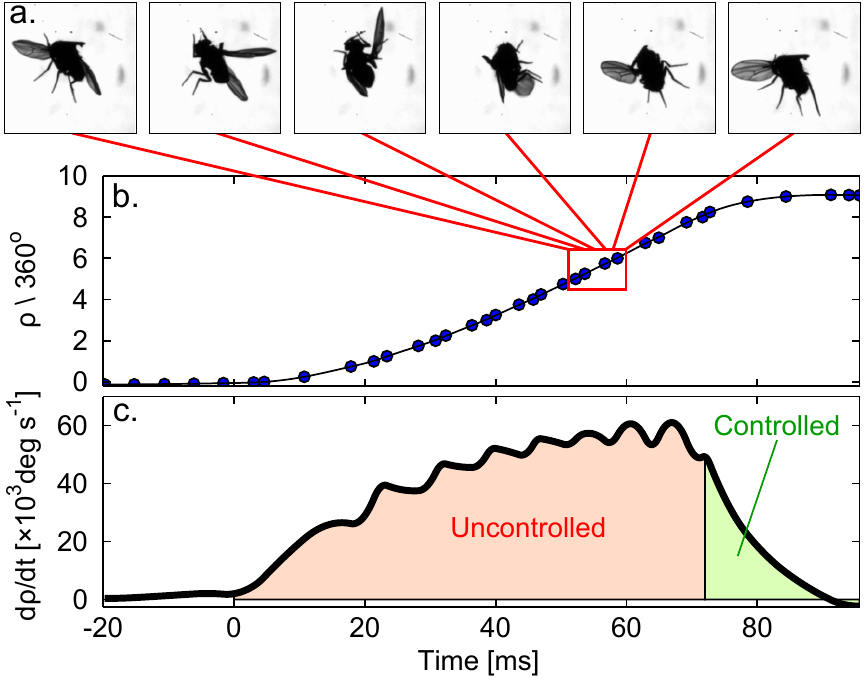}
\caption{\label{Fig4:Extreme}Extreme perturbation. (a) Snapshots, taken $1.25\ms$ apart, of a fly being rotated at $\sim60,000\deg\persec$ by a pulse train of magnetic torques. (b) Roll angle versus time during the maneuver (blue circles), and its smoothing spline (black line). The perturbation was on between $t=0-72\ms$. Measurement errors are smaller than the symbol size. (c) Roll velocity versus time (black line). The fly was unable to control its flight during the perturbation (red shading) and actively corrected roll once the perturbation stopped (green shading). The fly leveled itself to $\rho=20\deg$ within $4$ wing-beats after the perturbation stopped, approaching its original roll orientation.}
\end{figure}

If the fly's roll controller were a linear PI (Eq. 1), the integral term should have accumulated a signal corresponding to a $3000\deg$ deflection. The resulting correction maneuver would require the fly to rotate $3000\deg$ in the opposite direction. Clearly flies circumvent this scenario by employing an ``anti-windup'' operation that prevents such accumulations from taking place. It is plausible that in such maneuvers the flies incorporate an additional sensory modality to determine the direction of gravity and do not rely on integration of angular velocity. Either way, the observed behavior is an example of a nonlinear feature for roll control during extreme perturbations. 

\emph{Summary and outlook:} Using impulsive torques to perturb flies in mid-air, we have investigated the mechanism flies use to control their body roll angle. To generate corrective torques, the flies apply a stroke-amplitude asymmetry that is effectively described by the output of a linear PI controller, possibly with integral ``anti-windup''. The flies respond to perturbations within a single wing-beat, or $5\ms$, placing this response among the fastest reflexes in the animal kingdom, comparable to the head-roll compensation reflex in blowflies ($5-7.5\ms$) \cite{HengstenbergRoll1986} and to the startle response in teleost fish ($5-10\ms$) \cite{EatonJEB1977, EatonBook1984}. The roll response is twice as fast as the escape response of cockroaches ($11-18\ms$) \cite{CamhiCockroaches1981, JindrichFullJEB2002}, and five times faster than visual startle response in flies ($25\ms$) \cite{FotowatEscapeResponse2009}. Moreover, the roll response time in flies is much faster than their response time to yaw and pitch perturbations \cite{RistrophPNAS2010, RistrophPitchInterface2013}, which highlights the relative importance of roll control. The flies correct back to $10\%$ of the maximum roll deflection within $6.8\pm1.6$ wing-beats, comparable to their visual response time, indicating the fly's compound eyes are not used throughout the correction maneuver.

An open problem arising from our study is understanding the structure of the overarching controller for all the body angles. For example, the perturbing torques in our experiment also induced secondary deflections along yaw and pitch that were often left partially uncorrected (SI). Thus, the fly's controller may prioritize roll while compromising other angles. To map out the full controller it will be necessary to measure the insect's response to sophisticated perturbations along multiple axes, whose direction, amplitude and timing are individually controlled. Such measurements will further reveal the strategies insects use to manage their actuation resources and achieve the grace and performance of their flight.

\begin{acknowledgments}
\emph{Acknowledgments:} T.B. was supported by the Cross Disciplinary Postdoctoral Fellowship of the Human Frontier Science Program. In addition the work was supported in part by an NSF-CBET award number 0933332 and an ARO award number 61651-EG. We thank Andy Ruina, Brian Leahy, Sagi Levy, Cole Gilbert, Ron Hoy, Paul Shamble, Jesse Goldberg, Sarah Iams, Leif Ristroph, Simon Walker, Svetlana Morozova, Noah Cowan, Brandon Hencey, Jen Grenier and Andrew Clark. Finally, T.B. would like to thank Anat B. Gafen for her invaluable support.
\end{acknowledgments}

\bibliography{insectFlight}

\begin{thebibliography}{45}%
\makeatletter
\providecommand \@ifxundefined [1]{%
 \@ifx{#1\undefined}
}%
\providecommand \@ifnum [1]{%
 \ifnum #1\expandafter \@firstoftwo
 \else \expandafter \@secondoftwo
 \fi
}%
\providecommand \@ifx [1]{%
 \ifx #1\expandafter \@firstoftwo
 \else \expandafter \@secondoftwo
 \fi
}%
\providecommand \natexlab [1]{#1}%
\providecommand \enquote  [1]{``#1''}%
\providecommand \bibnamefont  [1]{#1}%
\providecommand \bibfnamefont [1]{#1}%
\providecommand \citenamefont [1]{#1}%
\providecommand \href@noop [0]{\@secondoftwo}%
\providecommand \href [0]{\begingroup \@sanitize@url \@href}%
\providecommand \@href[1]{\@@startlink{#1}\@@href}%
\providecommand \@@href[1]{\endgroup#1\@@endlink}%
\providecommand \@sanitize@url [0]{\catcode `\\12\catcode `\$12\catcode
  `\&12\catcode `\#12\catcode `\^12\catcode `\_12\catcode `\%12\relax}%
\providecommand \@@startlink[1]{}%
\providecommand \@@endlink[0]{}%
\providecommand \url  [0]{\begingroup\@sanitize@url \@url }%
\providecommand \@url [1]{\endgroup\@href {#1}{\urlprefix }}%
\providecommand \urlprefix  [0]{URL }%
\providecommand \Eprint [0]{\href }%
\providecommand \doibase [0]{http://dx.doi.org/}%
\providecommand \selectlanguage [0]{\@gobble}%
\providecommand \bibinfo  [0]{\@secondoftwo}%
\providecommand \bibfield  [0]{\@secondoftwo}%
\providecommand \translation [1]{[#1]}%
\providecommand \BibitemOpen [0]{}%
\providecommand \bibitemStop [0]{}%
\providecommand \bibitemNoStop [0]{.\EOS\space}%
\providecommand \EOS [0]{\spacefactor3000\relax}%
\providecommand \BibitemShut  [1]{\csname bibitem#1\endcsname}%
\let\auto@bib@innerbib\@empty
\bibitem [{\citenamefont {Combes}\ and\ \citenamefont
  {Dudley}(2009)}]{CombesTurbulencePNAS2009}%
  \BibitemOpen
  \bibfield  {author} {\bibinfo {author} {\bibfnamefont {S.~A.}\ \bibnamefont
  {Combes}}\ and\ \bibinfo {author} {\bibfnamefont {R.}~\bibnamefont
  {Dudley}},\ }\href@noop {} {\bibfield  {journal} {\bibinfo  {journal}
  {Proceedings of the National Academy of Sciences}\ }\textbf {\bibinfo
  {volume} {106}},\ \bibinfo {pages} {9105} (\bibinfo {year}
  {2009})}\BibitemShut {NoStop}%
\bibitem [{\citenamefont {Ravi}\ \emph {et~al.}(2013)\citenamefont {Ravi},
  \citenamefont {Crall}, \citenamefont {Fisher},\ and\ \citenamefont
  {Combes}}]{RaviCombesJEB2013}%
  \BibitemOpen
  \bibfield  {author} {\bibinfo {author} {\bibfnamefont {S.}~\bibnamefont
  {Ravi}}, \bibinfo {author} {\bibfnamefont {J.~D.}\ \bibnamefont {Crall}},
  \bibinfo {author} {\bibfnamefont {A.}~\bibnamefont {Fisher}}, \ and\ \bibinfo
  {author} {\bibfnamefont {S.~A.}\ \bibnamefont {Combes}},\ }\href@noop {}
  {\bibfield  {journal} {\bibinfo  {journal} {The Journal of experimental
  biology}\ }\textbf {\bibinfo {volume} {216}},\ \bibinfo {pages} {4299}
  (\bibinfo {year} {2013})}\BibitemShut {NoStop}%
\bibitem [{\citenamefont {Ortega-Jimenez}\ \emph {et~al.}(2013)\citenamefont
  {Ortega-Jimenez}, \citenamefont {Greeter}, \citenamefont {Mittal},\ and\
  \citenamefont {Hedrick}}]{OrtegaHedrickJEB2013}%
  \BibitemOpen
  \bibfield  {author} {\bibinfo {author} {\bibfnamefont {V.~M.}\ \bibnamefont
  {Ortega-Jimenez}}, \bibinfo {author} {\bibfnamefont {J.~S.}\ \bibnamefont
  {Greeter}}, \bibinfo {author} {\bibfnamefont {R.}~\bibnamefont {Mittal}}, \
  and\ \bibinfo {author} {\bibfnamefont {T.~L.}\ \bibnamefont {Hedrick}},\
  }\href@noop {} {\bibfield  {journal} {\bibinfo  {journal} {The Journal of
  experimental biology}\ ,\ \bibinfo {pages} {jeb}} (\bibinfo {year}
  {2013})}\BibitemShut {NoStop}%
\bibitem [{\citenamefont {Vance}\ \emph {et~al.}(2013)\citenamefont {Vance},
  \citenamefont {Faruque},\ and\ \citenamefont {Humbert}}]{VanceHumbert2013}%
  \BibitemOpen
  \bibfield  {author} {\bibinfo {author} {\bibfnamefont {J.}~\bibnamefont
  {Vance}}, \bibinfo {author} {\bibfnamefont {I.}~\bibnamefont {Faruque}}, \
  and\ \bibinfo {author} {\bibfnamefont {J.}~\bibnamefont {Humbert}},\
  }\href@noop {} {\bibfield  {journal} {\bibinfo  {journal} {Bioinspiration \&
  biomimetics}\ }\textbf {\bibinfo {volume} {8}},\ \bibinfo {pages} {016004}
  (\bibinfo {year} {2013})}\BibitemShut {NoStop}%
\bibitem [{\citenamefont {Sane}(2003)}]{SaneReview2003}%
  \BibitemOpen
  \bibfield  {author} {\bibinfo {author} {\bibfnamefont {S.~P.}\ \bibnamefont
  {Sane}},\ }\href@noop {} {\bibfield  {journal} {\bibinfo  {journal} {The
  Journal of experimental biology}\ }\textbf {\bibinfo {volume} {206}},\
  \bibinfo {pages} {4191} (\bibinfo {year} {2003})}\BibitemShut {NoStop}%
\bibitem [{\citenamefont {Wang}(2005)}]{WangReview2005}%
  \BibitemOpen
  \bibfield  {author} {\bibinfo {author} {\bibfnamefont {Z.~J.}\ \bibnamefont
  {Wang}},\ }\href@noop {} {\bibfield  {journal} {\bibinfo  {journal} {Annu.
  Rev. Fluid Mech.}\ }\textbf {\bibinfo {volume} {37}},\ \bibinfo {pages} {183}
  (\bibinfo {year} {2005})}\BibitemShut {NoStop}%
\bibitem [{\citenamefont {Taylor}\ and\ \citenamefont
  {Thomas}(2003)}]{TaylorThomasJEB2003}%
  \BibitemOpen
  \bibfield  {author} {\bibinfo {author} {\bibfnamefont {G.~K.}\ \bibnamefont
  {Taylor}}\ and\ \bibinfo {author} {\bibfnamefont {A.~L.}\ \bibnamefont
  {Thomas}},\ }\href@noop {} {\bibfield  {journal} {\bibinfo  {journal}
  {Journal of Experimental Biology}\ }\textbf {\bibinfo {volume} {206}},\
  \bibinfo {pages} {2803} (\bibinfo {year} {2003})}\BibitemShut {NoStop}%
\bibitem [{\citenamefont {Sun}\ and\ \citenamefont
  {Xiong}(2005)}]{SunBumblebeeJEB2005}%
  \BibitemOpen
  \bibfield  {author} {\bibinfo {author} {\bibfnamefont {M.}~\bibnamefont
  {Sun}}\ and\ \bibinfo {author} {\bibfnamefont {Y.}~\bibnamefont {Xiong}},\
  }\href@noop {} {\bibfield  {journal} {\bibinfo  {journal} {The Journal of
  experimental biology}\ }\textbf {\bibinfo {volume} {208}},\ \bibinfo {pages}
  {447} (\bibinfo {year} {2005})}\BibitemShut {NoStop}%
\bibitem [{\citenamefont {Taylor}\ and\ \citenamefont
  {{\.Z}bikowski}(2005)}]{TaylorZibkowski2005}%
  \BibitemOpen
  \bibfield  {author} {\bibinfo {author} {\bibfnamefont {G.~K.}\ \bibnamefont
  {Taylor}}\ and\ \bibinfo {author} {\bibfnamefont {R.}~\bibnamefont
  {{\.Z}bikowski}},\ }\href@noop {} {\bibfield  {journal} {\bibinfo  {journal}
  {Journal of The Royal Society Interface}\ }\textbf {\bibinfo {volume} {2}},\
  \bibinfo {pages} {197} (\bibinfo {year} {2005})}\BibitemShut {NoStop}%
\bibitem [{\citenamefont {Sun}\ \emph {et~al.}(2007)\citenamefont {Sun},
  \citenamefont {Wang},\ and\ \citenamefont {Xiong}}]{SunDynamicStability2007}%
  \BibitemOpen
  \bibfield  {author} {\bibinfo {author} {\bibfnamefont {M.}~\bibnamefont
  {Sun}}, \bibinfo {author} {\bibfnamefont {J.}~\bibnamefont {Wang}}, \ and\
  \bibinfo {author} {\bibfnamefont {Y.}~\bibnamefont {Xiong}},\ }\href@noop {}
  {\bibfield  {journal} {\bibinfo  {journal} {Acta Mechanica Sinica}\ }\textbf
  {\bibinfo {volume} {23}},\ \bibinfo {pages} {231} (\bibinfo {year}
  {2007})}\BibitemShut {NoStop}%
\bibitem [{\citenamefont {Liu}\ \emph {et~al.}(2010)\citenamefont {Liu},
  \citenamefont {Nakata}, \citenamefont {Gao}, \citenamefont {Maeda},
  \citenamefont {Aono},\ and\ \citenamefont {Shyy}}]{LiuCFD2010}%
  \BibitemOpen
  \bibfield  {author} {\bibinfo {author} {\bibfnamefont {H.}~\bibnamefont
  {Liu}}, \bibinfo {author} {\bibfnamefont {T.}~\bibnamefont {Nakata}},
  \bibinfo {author} {\bibfnamefont {N.}~\bibnamefont {Gao}}, \bibinfo {author}
  {\bibfnamefont {M.}~\bibnamefont {Maeda}}, \bibinfo {author} {\bibfnamefont
  {H.}~\bibnamefont {Aono}}, \ and\ \bibinfo {author} {\bibfnamefont
  {W.}~\bibnamefont {Shyy}},\ }\href@noop {} {\bibfield  {journal} {\bibinfo
  {journal} {Acta Mechanica Sinica}\ }\textbf {\bibinfo {volume} {26}},\
  \bibinfo {pages} {863} (\bibinfo {year} {2010})}\BibitemShut {NoStop}%
\bibitem [{\citenamefont {Faruque}\ and\ \citenamefont
  {Humbert}(2010)}]{FaruqueHumbert2010a}%
  \BibitemOpen
  \bibfield  {author} {\bibinfo {author} {\bibfnamefont {I.}~\bibnamefont
  {Faruque}}\ and\ \bibinfo {author} {\bibfnamefont {J.~S.}\ \bibnamefont
  {Humbert}},\ }\href@noop {} {\bibfield  {journal} {\bibinfo  {journal}
  {Journal of theoretical biology}\ }\textbf {\bibinfo {volume} {264}},\
  \bibinfo {pages} {538} (\bibinfo {year} {2010})}\BibitemShut {NoStop}%
\bibitem [{\citenamefont {Ristroph}\ \emph {et~al.}(2013)\citenamefont
  {Ristroph}, \citenamefont {Ristroph}, \citenamefont {Morozova}, \citenamefont
  {Bergou}, \citenamefont {Chang}, \citenamefont {Guckenheimer}, \citenamefont
  {Wang},\ and\ \citenamefont {Cohen}}]{RistrophPitchInterface2013}%
  \BibitemOpen
  \bibfield  {author} {\bibinfo {author} {\bibfnamefont {L.}~\bibnamefont
  {Ristroph}}, \bibinfo {author} {\bibfnamefont {G.}~\bibnamefont {Ristroph}},
  \bibinfo {author} {\bibfnamefont {S.}~\bibnamefont {Morozova}}, \bibinfo
  {author} {\bibfnamefont {A.~J.}\ \bibnamefont {Bergou}}, \bibinfo {author}
  {\bibfnamefont {S.}~\bibnamefont {Chang}}, \bibinfo {author} {\bibfnamefont
  {J.}~\bibnamefont {Guckenheimer}}, \bibinfo {author} {\bibfnamefont {Z.~J.}\
  \bibnamefont {Wang}}, \ and\ \bibinfo {author} {\bibfnamefont
  {I.}~\bibnamefont {Cohen}},\ }\href@noop {} {\bibfield  {journal} {\bibinfo
  {journal} {Journal of The Royal Society Interface}\ }\textbf {\bibinfo
  {volume} {10}} (\bibinfo {year} {2013})}\BibitemShut {NoStop}%
\bibitem [{\citenamefont {Gao}\ \emph {et~al.}(2011)\citenamefont {Gao},
  \citenamefont {Aono},\ and\ \citenamefont {Liu}}]{GaoCFD2011}%
  \BibitemOpen
  \bibfield  {author} {\bibinfo {author} {\bibfnamefont {N.}~\bibnamefont
  {Gao}}, \bibinfo {author} {\bibfnamefont {H.}~\bibnamefont {Aono}}, \ and\
  \bibinfo {author} {\bibfnamefont {H.}~\bibnamefont {Liu}},\ }\href@noop {}
  {\bibfield  {journal} {\bibinfo  {journal} {Journal of Theoretical Biology}\
  }\textbf {\bibinfo {volume} {270}},\ \bibinfo {pages} {98} (\bibinfo {year}
  {2011})}\BibitemShut {NoStop}%
\bibitem [{\citenamefont {Liang}\ and\ \citenamefont
  {Sun}(2013)}]{LiangSunInterface2013}%
  \BibitemOpen
  \bibfield  {author} {\bibinfo {author} {\bibfnamefont {B.}~\bibnamefont
  {Liang}}\ and\ \bibinfo {author} {\bibfnamefont {M.}~\bibnamefont {Sun}},\
  }\href@noop {} {\bibfield  {journal} {\bibinfo  {journal} {Journal of The
  Royal Society Interface}\ }\textbf {\bibinfo {volume} {10}} (\bibinfo {year}
  {2013})}\BibitemShut {NoStop}%
\bibitem [{\citenamefont {Xu}\ and\ \citenamefont {Sun}(2013)}]{XuSunCFD2013}%
  \BibitemOpen
  \bibfield  {author} {\bibinfo {author} {\bibfnamefont {N.}~\bibnamefont
  {Xu}}\ and\ \bibinfo {author} {\bibfnamefont {M.}~\bibnamefont {Sun}},\
  }\href@noop {} {\bibfield  {journal} {\bibinfo  {journal} {Journal of
  theoretical biology}\ }\textbf {\bibinfo {volume} {319}},\ \bibinfo {pages}
  {102} (\bibinfo {year} {2013})}\BibitemShut {NoStop}%
\bibitem [{\citenamefont {Zhang}\ and\ \citenamefont
  {Sun}(2010)}]{ZhangSunLateral2010}%
  \BibitemOpen
  \bibfield  {author} {\bibinfo {author} {\bibfnamefont {Y.}~\bibnamefont
  {Zhang}}\ and\ \bibinfo {author} {\bibfnamefont {M.}~\bibnamefont {Sun}},\
  }\href@noop {} {\bibfield  {journal} {\bibinfo  {journal} {Acta Mechanica
  Sinica}\ }\textbf {\bibinfo {volume} {26}},\ \bibinfo {pages} {175} (\bibinfo
  {year} {2010})}\BibitemShut {NoStop}%
\bibitem [{\citenamefont {Nalbach}(1994)}]{NalbachNonOrthogonal1994}%
  \BibitemOpen
  \bibfield  {author} {\bibinfo {author} {\bibfnamefont {G.}~\bibnamefont
  {Nalbach}},\ }\href@noop {} {\bibfield  {journal} {\bibinfo  {journal}
  {Neuroscience}\ }\textbf {\bibinfo {volume} {61}},\ \bibinfo {pages} {149}
  (\bibinfo {year} {1994})}\BibitemShut {NoStop}%
\bibitem [{\citenamefont {Dickinson}(1999)}]{DickinsonHaltere1999}%
  \BibitemOpen
  \bibfield  {author} {\bibinfo {author} {\bibfnamefont {M.~H.}\ \bibnamefont
  {Dickinson}},\ }\href@noop {} {\bibfield  {journal} {\bibinfo  {journal}
  {Philosophical Transactions of the Royal Society of London.Series B:
  Biological Sciences}\ }\textbf {\bibinfo {volume} {354}},\ \bibinfo {pages}
  {903} (\bibinfo {year} {1999})}\BibitemShut {NoStop}%
\bibitem [{\citenamefont {Sherman}\ and\ \citenamefont
  {Dickinson}(2004)}]{ShermanDickinsonJEB2003}%
  \BibitemOpen
  \bibfield  {author} {\bibinfo {author} {\bibfnamefont {A.}~\bibnamefont
  {Sherman}}\ and\ \bibinfo {author} {\bibfnamefont {M.~H.}\ \bibnamefont
  {Dickinson}},\ }\href@noop {} {\bibfield  {journal} {\bibinfo  {journal}
  {Journal of experimental biology}\ }\textbf {\bibinfo {volume} {207}},\
  \bibinfo {pages} {133} (\bibinfo {year} {2004})}\BibitemShut {NoStop}%
\bibitem [{\citenamefont {Sherman}\ and\ \citenamefont
  {Dickinson}(2003)}]{ShermanDickinsonJEB2004}%
  \BibitemOpen
  \bibfield  {author} {\bibinfo {author} {\bibfnamefont {A.}~\bibnamefont
  {Sherman}}\ and\ \bibinfo {author} {\bibfnamefont {M.~H.}\ \bibnamefont
  {Dickinson}},\ }\href@noop {} {\bibfield  {journal} {\bibinfo  {journal}
  {Journal of Experimental Biology}\ }\textbf {\bibinfo {volume} {206}},\
  \bibinfo {pages} {295} (\bibinfo {year} {2003})}\BibitemShut {NoStop}%
\bibitem [{\citenamefont {Sugiura}\ and\ \citenamefont
  {Dickinson}(2009)}]{SugiuraDickinsonPlotOne2009}%
  \BibitemOpen
  \bibfield  {author} {\bibinfo {author} {\bibfnamefont {H.}~\bibnamefont
  {Sugiura}}\ and\ \bibinfo {author} {\bibfnamefont {M.~H.}\ \bibnamefont
  {Dickinson}},\ }\href@noop {} {\bibfield  {journal} {\bibinfo  {journal}
  {PloS one}\ }\textbf {\bibinfo {volume} {4}},\ \bibinfo {pages} {e4883}
  (\bibinfo {year} {2009})}\BibitemShut {NoStop}%
\bibitem [{\citenamefont {Hengstenberg}\ \emph {et~al.}(1986)\citenamefont
  {Hengstenberg}, \citenamefont {Sandeman},\ and\ \citenamefont
  {Hengstenberg}}]{HengstenbergRoll1986}%
  \BibitemOpen
  \bibfield  {author} {\bibinfo {author} {\bibfnamefont {R.}~\bibnamefont
  {Hengstenberg}}, \bibinfo {author} {\bibfnamefont {D.}~\bibnamefont
  {Sandeman}}, \ and\ \bibinfo {author} {\bibfnamefont {B.}~\bibnamefont
  {Hengstenberg}},\ }\href@noop {} {\bibfield  {journal} {\bibinfo  {journal}
  {Proceedings of the Royal society of London. Series B. Biological sciences}\
  }\textbf {\bibinfo {volume} {227}},\ \bibinfo {pages} {455} (\bibinfo {year}
  {1986})}\BibitemShut {NoStop}%
\bibitem [{\citenamefont {Srinivasan}(1977)}]{SrinivasanVisual1977}%
  \BibitemOpen
  \bibfield  {author} {\bibinfo {author} {\bibfnamefont {M.~V.}\ \bibnamefont
  {Srinivasan}},\ }\href@noop {} {\bibfield  {journal} {\bibinfo  {journal}
  {Journal of comparative physiology}\ }\textbf {\bibinfo {volume} {119}},\
  \bibinfo {pages} {1} (\bibinfo {year} {1977})}\BibitemShut {NoStop}%
\bibitem [{\citenamefont {Waldmann}\ and\ \citenamefont
  {Zarnack}(1988)}]{WaldmannLocust1988}%
  \BibitemOpen
  \bibfield  {author} {\bibinfo {author} {\bibfnamefont {B.}~\bibnamefont
  {Waldmann}}\ and\ \bibinfo {author} {\bibfnamefont {W.}~\bibnamefont
  {Zarnack}},\ }\href@noop {} {\bibfield  {journal} {\bibinfo  {journal}
  {Biological cybernetics}\ }\textbf {\bibinfo {volume} {59}},\ \bibinfo
  {pages} {325} (\bibinfo {year} {1988})}\BibitemShut {NoStop}%
\bibitem [{\citenamefont {Lehmann}\ \emph {et~al.}(1998)\citenamefont {Lehmann}
  \emph {et~al.}}]{LehmannDickinson1998}%
  \BibitemOpen
  \bibfield  {author} {\bibinfo {author} {\bibfnamefont {F.-O.}\ \bibnamefont
  {Lehmann}} \emph {et~al.},\ }\href@noop {} {\bibfield  {journal} {\bibinfo
  {journal} {The Journal of experimental biology}\ }\textbf {\bibinfo {volume}
  {201}},\ \bibinfo {pages} {385} (\bibinfo {year} {1998})}\BibitemShut
  {NoStop}%
\bibitem [{\citenamefont {Zanker}(1990)}]{ZankerIIIControl1990}%
  \BibitemOpen
  \bibfield  {author} {\bibinfo {author} {\bibfnamefont {J.}~\bibnamefont
  {Zanker}},\ }\href@noop {} {\bibfield  {journal} {\bibinfo  {journal}
  {Philosophical Transactions of the Royal Society of London. B, Biological
  Sciences}\ }\textbf {\bibinfo {volume} {327}},\ \bibinfo {pages} {43}
  (\bibinfo {year} {1990})}\BibitemShut {NoStop}%
\bibitem [{\citenamefont {Windsor}\ \emph {et~al.}(2014)\citenamefont
  {Windsor}, \citenamefont {Bomphrey},\ and\ \citenamefont
  {Taylor}}]{WindsorInterface2014}%
  \BibitemOpen
  \bibfield  {author} {\bibinfo {author} {\bibfnamefont {S.~P.}\ \bibnamefont
  {Windsor}}, \bibinfo {author} {\bibfnamefont {R.~J.}\ \bibnamefont
  {Bomphrey}}, \ and\ \bibinfo {author} {\bibfnamefont {G.~K.}\ \bibnamefont
  {Taylor}},\ }\href@noop {} {\bibfield  {journal} {\bibinfo  {journal}
  {Journal of The Royal Society Interface}\ }\textbf {\bibinfo {volume} {11}},\
  \bibinfo {pages} {20130921} (\bibinfo {year} {2014})}\BibitemShut {NoStop}%
\bibitem [{\citenamefont {Bender}\ and\ \citenamefont
  {Dickinson}(2006)}]{BenderDickinsonJEB2006}%
  \BibitemOpen
  \bibfield  {author} {\bibinfo {author} {\bibfnamefont {J.~A.}\ \bibnamefont
  {Bender}}\ and\ \bibinfo {author} {\bibfnamefont {M.~H.}\ \bibnamefont
  {Dickinson}},\ }\href@noop {} {\bibfield  {journal} {\bibinfo  {journal} {The
  Journal of experimental biology}\ }\textbf {\bibinfo {volume} {209}},\
  \bibinfo {pages} {3170} (\bibinfo {year} {2006})}\BibitemShut {NoStop}%
\bibitem [{\citenamefont {Ristroph}\ \emph {et~al.}(2010)\citenamefont
  {Ristroph}, \citenamefont {Bergou}, \citenamefont {Ristroph}, \citenamefont
  {Coumes}, \citenamefont {Berman}, \citenamefont {Guckenheimer}, \citenamefont
  {Wang},\ and\ \citenamefont {Cohen}}]{RistrophPNAS2010}%
  \BibitemOpen
  \bibfield  {author} {\bibinfo {author} {\bibfnamefont {L.}~\bibnamefont
  {Ristroph}}, \bibinfo {author} {\bibfnamefont {A.~J.}\ \bibnamefont
  {Bergou}}, \bibinfo {author} {\bibfnamefont {G.}~\bibnamefont {Ristroph}},
  \bibinfo {author} {\bibfnamefont {K.}~\bibnamefont {Coumes}}, \bibinfo
  {author} {\bibfnamefont {G.~J.}\ \bibnamefont {Berman}}, \bibinfo {author}
  {\bibfnamefont {J.}~\bibnamefont {Guckenheimer}}, \bibinfo {author}
  {\bibfnamefont {Z.~J.}\ \bibnamefont {Wang}}, \ and\ \bibinfo {author}
  {\bibfnamefont {I.}~\bibnamefont {Cohen}},\ }\href@noop {} {\bibfield
  {journal} {\bibinfo  {journal} {Proceedings of the National Academy of
  Sciences}\ }\textbf {\bibinfo {volume} {107}},\ \bibinfo {pages} {4820}
  (\bibinfo {year} {2010})}\BibitemShut {NoStop}%
\bibitem [{\citenamefont {Ristroph}\ \emph {et~al.}(2009)\citenamefont
  {Ristroph}, \citenamefont {Berman}, \citenamefont {Bergou}, \citenamefont
  {Wang},\ and\ \citenamefont {Cohen}}]{RistrophHRMT_JEB2009}%
  \BibitemOpen
  \bibfield  {author} {\bibinfo {author} {\bibfnamefont {L.}~\bibnamefont
  {Ristroph}}, \bibinfo {author} {\bibfnamefont {G.~J.}\ \bibnamefont
  {Berman}}, \bibinfo {author} {\bibfnamefont {A.~J.}\ \bibnamefont {Bergou}},
  \bibinfo {author} {\bibfnamefont {Z.~J.}\ \bibnamefont {Wang}}, \ and\
  \bibinfo {author} {\bibfnamefont {I.}~\bibnamefont {Cohen}},\ }\href@noop {}
  {\bibfield  {journal} {\bibinfo  {journal} {Journal of Experimental Biology}\
  }\textbf {\bibinfo {volume} {212}},\ \bibinfo {pages} {1324} (\bibinfo {year}
  {2009})}\BibitemShut {NoStop}%
\bibitem [{\citenamefont {Fotowat}\ \emph {et~al.}(2009)\citenamefont
  {Fotowat}, \citenamefont {Fayyazuddin}, \citenamefont {Bellen},\ and\
  \citenamefont {Gabbiani}}]{FotowatEscapeResponse2009}%
  \BibitemOpen
  \bibfield  {author} {\bibinfo {author} {\bibfnamefont {H.}~\bibnamefont
  {Fotowat}}, \bibinfo {author} {\bibfnamefont {A.}~\bibnamefont
  {Fayyazuddin}}, \bibinfo {author} {\bibfnamefont {H.~J.}\ \bibnamefont
  {Bellen}}, \ and\ \bibinfo {author} {\bibfnamefont {F.}~\bibnamefont
  {Gabbiani}},\ }\href@noop {} {\bibfield  {journal} {\bibinfo  {journal}
  {Journal of neurophysiology}\ }\textbf {\bibinfo {volume} {102}},\ \bibinfo
  {pages} {875} (\bibinfo {year} {2009})}\BibitemShut {NoStop}%
\bibitem [{\citenamefont {Land}\ and\ \citenamefont
  {Collett}(1974)}]{LandCollet1974}%
  \BibitemOpen
  \bibfield  {author} {\bibinfo {author} {\bibfnamefont {M.~F.}\ \bibnamefont
  {Land}}\ and\ \bibinfo {author} {\bibfnamefont {T.}~\bibnamefont {Collett}},\
  }\href@noop {} {\bibfield  {journal} {\bibinfo  {journal} {Journal of
  Comparative Physiology}\ }\textbf {\bibinfo {volume} {89}},\ \bibinfo {pages}
  {331} (\bibinfo {year} {1974})}\BibitemShut {NoStop}%
\bibitem [{\citenamefont {Eaton}(1984)}]{EatonBook1984}%
  \BibitemOpen
  \bibfield  {author} {\bibinfo {author} {\bibfnamefont {R.~C.}\ \bibnamefont
  {Eaton}},\ }\href@noop {} {\emph {\bibinfo {title} {Neural mechanisms of
  startle behavior}}}\ (\bibinfo  {publisher} {Springer},\ \bibinfo {year}
  {1984})\BibitemShut {NoStop}%
\bibitem [{\citenamefont {Goodman}(1960)}]{goodman1960landing}%
  \BibitemOpen
  \bibfield  {author} {\bibinfo {author} {\bibfnamefont {L.~J.}\ \bibnamefont
  {Goodman}},\ }\href@noop {} {\bibfield  {journal} {\bibinfo  {journal}
  {Journal of Experimental Biology}\ }\textbf {\bibinfo {volume} {37}},\
  \bibinfo {pages} {854} (\bibinfo {year} {1960})}\BibitemShut {NoStop}%
\bibitem [{\citenamefont {Borst}(1986)}]{borst1986time}%
  \BibitemOpen
  \bibfield  {author} {\bibinfo {author} {\bibfnamefont {A.}~\bibnamefont
  {Borst}},\ }\href@noop {} {\bibfield  {journal} {\bibinfo  {journal}
  {Biological cybernetics}\ }\textbf {\bibinfo {volume} {54}},\ \bibinfo
  {pages} {379} (\bibinfo {year} {1986})}\BibitemShut {NoStop}%
\bibitem [{\citenamefont {Van~Breugel}\ and\ \citenamefont
  {Dickinson}(2012)}]{vanBreugel2012visual}%
  \BibitemOpen
  \bibfield  {author} {\bibinfo {author} {\bibfnamefont {F.}~\bibnamefont
  {Van~Breugel}}\ and\ \bibinfo {author} {\bibfnamefont {M.~H.}\ \bibnamefont
  {Dickinson}},\ }\href@noop {} {\bibfield  {journal} {\bibinfo  {journal} {The
  Journal of experimental biology}\ }\textbf {\bibinfo {volume} {215}},\
  \bibinfo {pages} {1783} (\bibinfo {year} {2012})}\BibitemShut {NoStop}%
\bibitem [{\citenamefont {Dickinson}\ \emph {et~al.}(1999)\citenamefont
  {Dickinson}, \citenamefont {Lehmann},\ and\ \citenamefont
  {Sane}}]{DickinsonScience1999}%
  \BibitemOpen
  \bibfield  {author} {\bibinfo {author} {\bibfnamefont {M.~H.}\ \bibnamefont
  {Dickinson}}, \bibinfo {author} {\bibfnamefont {F.~O.}\ \bibnamefont
  {Lehmann}}, \ and\ \bibinfo {author} {\bibfnamefont {S.~P.}\ \bibnamefont
  {Sane}},\ }\href@noop {} {\bibfield  {journal} {\bibinfo  {journal}
  {Science}\ }\textbf {\bibinfo {volume} {284}},\ \bibinfo {pages} {1954}
  (\bibinfo {year} {1999})}\BibitemShut {NoStop}%
\bibitem [{\citenamefont {Sane}\ and\ \citenamefont
  {Dickinson}(2002)}]{SaneDickinsonJEB2002}%
  \BibitemOpen
  \bibfield  {author} {\bibinfo {author} {\bibfnamefont {S.~P.}\ \bibnamefont
  {Sane}}\ and\ \bibinfo {author} {\bibfnamefont {M.~H.}\ \bibnamefont
  {Dickinson}},\ }\href@noop {} {\bibfield  {journal} {\bibinfo  {journal}
  {Journal of Experimental Biology}\ }\textbf {\bibinfo {volume} {205}},\
  \bibinfo {pages} {1087} (\bibinfo {year} {2002})}\BibitemShut {NoStop}%
\bibitem [{\citenamefont {Berman}\ and\ \citenamefont
  {Wang}(2007)}]{BermanWangJFM2007}%
  \BibitemOpen
  \bibfield  {author} {\bibinfo {author} {\bibfnamefont {G.~J.}\ \bibnamefont
  {Berman}}\ and\ \bibinfo {author} {\bibfnamefont {Z.~J.}\ \bibnamefont
  {Wang}},\ }\href@noop {} {\bibfield  {journal} {\bibinfo  {journal} {Journal
  of Fluid Mechanics}\ }\textbf {\bibinfo {volume} {582}},\ \bibinfo {pages}
  {153} (\bibinfo {year} {2007})}\BibitemShut {NoStop}%
\bibitem [{\citenamefont {Pringle}(1948)}]{Pringle1948}%
  \BibitemOpen
  \bibfield  {author} {\bibinfo {author} {\bibfnamefont {J.}~\bibnamefont
  {Pringle}},\ }\href@noop {} {\bibfield  {journal} {\bibinfo  {journal} {Royal
  Society of London Philosophical Transactions Series B}\ }\textbf {\bibinfo
  {volume} {233}},\ \bibinfo {pages} {347} (\bibinfo {year}
  {1948})}\BibitemShut {NoStop}%
\bibitem [{\citenamefont {Nalbach}(1993)}]{NalbachHalteres1993}%
  \BibitemOpen
  \bibfield  {author} {\bibinfo {author} {\bibfnamefont {G.}~\bibnamefont
  {Nalbach}},\ }\href@noop {} {\bibfield  {journal} {\bibinfo  {journal}
  {Journal of Comparative Physiology A: Neuroethology, Sensory, Neural, and
  Behavioral Physiology}\ }\textbf {\bibinfo {volume} {173}},\ \bibinfo {pages}
  {293} (\bibinfo {year} {1993})}\BibitemShut {NoStop}%
\bibitem [{\citenamefont {Eaton}\ \emph {et~al.}(1977)\citenamefont {Eaton},
  \citenamefont {Bombardieri},\ and\ \citenamefont {Meyer}}]{EatonJEB1977}%
  \BibitemOpen
  \bibfield  {author} {\bibinfo {author} {\bibfnamefont {R.~C.}\ \bibnamefont
  {Eaton}}, \bibinfo {author} {\bibfnamefont {R.~A.}\ \bibnamefont
  {Bombardieri}}, \ and\ \bibinfo {author} {\bibfnamefont {D.~L.}\ \bibnamefont
  {Meyer}},\ }\href@noop {} {\bibfield  {journal} {\bibinfo  {journal} {Journal
  of Experimental Biology}\ }\textbf {\bibinfo {volume} {66}},\ \bibinfo
  {pages} {65} (\bibinfo {year} {1977})}\BibitemShut {NoStop}%
\bibitem [{\citenamefont {Camhi}\ and\ \citenamefont
  {Nolen}(1981)}]{CamhiCockroaches1981}%
  \BibitemOpen
  \bibfield  {author} {\bibinfo {author} {\bibfnamefont {J.~M.}\ \bibnamefont
  {Camhi}}\ and\ \bibinfo {author} {\bibfnamefont {T.~G.}\ \bibnamefont
  {Nolen}},\ }\href@noop {} {\bibfield  {journal} {\bibinfo  {journal} {Journal
  of comparative physiology}\ }\textbf {\bibinfo {volume} {142}},\ \bibinfo
  {pages} {339} (\bibinfo {year} {1981})}\BibitemShut {NoStop}%
\bibitem [{\citenamefont {Jindrich}\ and\ \citenamefont
  {Full}(2002)}]{JindrichFullJEB2002}%
  \BibitemOpen
  \bibfield  {author} {\bibinfo {author} {\bibfnamefont {D.~L.}\ \bibnamefont
  {Jindrich}}\ and\ \bibinfo {author} {\bibfnamefont {R.~J.}\ \bibnamefont
  {Full}},\ }\href@noop {} {\bibfield  {journal} {\bibinfo  {journal} {Journal
  of Experimental Biology}\ }\textbf {\bibinfo {volume} {205}},\ \bibinfo
  {pages} {2803} (\bibinfo {year} {2002})}\BibitemShut {NoStop}%
\end{thebibliography}%


\end{document}